\documentclass[12pt]{article}
\usepackage{amsfonts}
\usepackage{amssymb}
\usepackage{amsmath}
\usepackage{graphicx}
\usepackage{setspace}
\usepackage{amscd}
\newcommand{\bt}[1]{\textbf{#1}} 
\newtheorem{theorem}{Theorem}
\newtheorem{acknowledgement}{Acknowledgement}

\newtheorem{corollary}{Corollary}

\begin{document}
\title{A supermartingale
argument for characterizing the Functional Hill process weak law for small
parameters}
\author{{Dr. A. M. Fall}\footnote{A. M. Fall (adjambarkafall@gmail.com) is a seasonal professor,
University Gaston Berger (UGB), Saint Louis, Senegal;}\and{G. S. Lo}\footnote{G. S. Lo
(gane-samb.lo@ugb.edu.sn) is a Professor, UGB, Saint Louis, Senegal; and African
University of Sciences and technolgy (AUST), Abuja, Nigeria;
and Associated to LSTA, Pierre et Marie Curie University, Paris, France;}\and {A. Adekpedjou}\footnote{A. Adekpedjou
(akima@mst.edu) is Associate Professor, Missouri S\&T, Rolla, MO, 65409-USA} \and{C. H. Ndiaye}\footnote{C. H. Ndiaye (ham111266@yahoo.fr) LMA, UCAD, Dakar, Senegal} 
}
\maketitle
\begin{abstract}
The paper deals with the asymptotic laws of functionals of standard exponential random variables. These classes of statistics are
closely related to estimators of the extreme value index when the underlying distribution function is in the Weibull domain of attraction. We use
techniques based on martingales theory to describe the non Gaussian asymptotic distribution of the aforementioned statistics.
We provide results of a simulation study as well  as statistical tests that may be of interest with the proposed results\\

\bigskip \noindent
\emph{\bt{Keywords:}} Supermartingale; Functional Hill process; Extreme
value theory
\end{abstract}
\onehalfspacing
\section{Introduction}
\label{sec1}
\label{subsec31}
\noindent Of interest in this manuscript are the asymptotic properties of functional stochastic processes
based on extreme values of independent and identically distributed (iid) random variables (\textit{rv})
$X_{1},X_{2},...$, wich are defined on the same probability space $(\Omega, \mathcal{P},\mathbb{P})$, whose common distribution function is $F$. Let $X_{1,n}\leq \cdots\leq X_{n,n}$ be the associated order statistics and $k(n)$ a
sequence of integers satisfying $1\leq k(n)<n$.\\

\noindent Since we are only interested by the upper tail of $F$, we may and do suppose that $X_i > 0$. Accordingly, we will have to use the log transform $Y=\log X$ to get iid rv's $Y_i=\log X_i$, $i \geq 1$, with common distribution function (\textit{df}) $G(x)=F(e^x)$ such that their order statistics satisfy $\{Y_{i,n}, 1\leq i \leq n   \}=\{\log X_{i,n}, 1\leq i \leq n \}$ for each $n \geq 1$. \\

\noindent Let $f(j)$ be a real and increasing function of the integer $j$ such that $f(0)=0$. The following empirical process, hereon called the functional Hill process,
\begin{equation}
T_n(f)=\sum_{j=1}^{k(n)}f(j)\left( \log X_{n-j+1,n}-\log X_{n-j,n}\right),
\label{hf1}
\end{equation}

\noindent was introduced by Deme et al. (2012). A generalization of
the Diop et al. statistic given by
\begin{equation}
\sum_{j=1}^{k(n)}j^{\tau }\left( \log X_{n-j+1,n}-\log X_{n-j,n}\right)
/k^{\tau}(n),  \label{hft}
\end{equation}

\noindent for $f(j)=j^{\tau }$, $\tau>0$, $j\geq 1$ and $f(0)=0$ is obtained when (\ref{hf1}) is divided by $f(k)$  (cf. Diop and Lo (1994) and Diop and Lo (2009)). The statistic (\ref{hft}) is a direct generalization of the classical Hill's estimator obtained for $\tau=1$. If $K$ is some Kernel function (cf. Goegebeur et al. (2010) and
Groeneboom et al. (2003)), these statistics are closely related to the Kernel-type estimators like the one due to Cs\"{o}rg\H{o} et al. (1985) and given by
\begin{equation}\normalsize
\left\{ \sum_{j=1}^{k(n)}jK(j/k(n)) \left( \log X_{n-j+1,n}-\log
X_{n-j,n}\right) /k(n)\right\}\big /\left\{
\sum_{j=1}^{k(n)}K(j/k(n))\right\}.  \label{cdm}
\end{equation}

\indent The statistics (\ref{hft}) and (\ref{cdm}) are generalizations of the Hill estimator corresponding to $K=1$ in (\ref{cdm}) and $\tau =1$ in (\ref{hft}) respectively, the later playing a crucial role in Extreme Value Theory (EVT). EVT is a theory whose main purpose pertains to the derivation of the asymptotic distribution of the maximum order statistics for a random sample $X_1,...,X_n$, usually denoted by $X_{n,n}$. The distribution of $X_{n,n}$ has applications in many areas such as actuarial science in the determination of the distributional properties of the largest claim, or in survival analysis to determine the asymptotic properties of the largest failure time of items subject to failure or censored data to name a few.\\

\noindent One says that the underlying distribution function  $F$ of the $X_i$'s is attracted to another distribution function $H$ if
for some sequences $(a_{n}>0)_{n\geq 1}$ and $(b_{n})_{n\geq 1}$, and for any real point of continuity $x$ of $H$, the following holds
\begin{equation}\label{attrac}
\lim_{n\rightarrow \infty }P\left(\frac{X_{n,n}-b_{n}}{a_{n}}\leq
x\right)=\lim_{n\rightarrow \infty }F^{n}(a_{n}x+b_{n})=H(x).
\end{equation}

\noindent If $H$ is non-degenerated, then it can be parameterized as $%
G_{\gamma }(x)=\exp [-(1+\gamma x)^{-1/\gamma }]I\{1+\gamma x>0\}$ with $\gamma\neq0$, where $I\{\cdot\}$ denotes the indicator function and $G_{0}(x)=\exp(-\exp(-x))$. Here, the function $G_{\gamma}(x)$ is called the Generalized Extreme Value (GEV)
distribution. If property (\ref{attrac}) holds, one says that $F$ is
in the domain of attraction of $H$, and we write $F\in D(H)$. The reader is referred to
de Haan and Ferreira (2006), Resnick (1987), Galambos (1985), Beirlant et al. (2004), and Embrechts
et al. (1997) for a modern account of EVT. Although the parameter $\gamma $ in the GEV distribution is
continuous, the three cases $\gamma <0$, $\gamma =0$ and $\gamma >0,$
named after the Weibull, Gumbel and Frechet domains of attraction respectively have different
distributional properties. However, in all three cases, the Hill statistic is used to
estimate the  extreme value index corresponding to $\tau=1$ in (\ref{hft}).\\

\noindent (1) In the case of $\gamma \geq 0$ for instance, (\ref{hft}) converges in probability to $\gamma$ as $n\rightarrow \infty$ and $k/n\rightarrow 0$.\\

\noindent (2), For $\gamma <0$, the upper endpoint of $G(x)=F(e^x)$ defined by $y_0=\sup \{x\in \Re, G(x)<1\}$ is finite, and it is related to the upper endpoint of $F$, $x_0=\sup \{x\in \Re, F(x)<1\}$ by $y_0=\log x_0$. If $G^{-1}$ stands for the generalized inverse function of $G$, then (\ref{hft}), when normalized by $y_{0}-G^{-1}(1-k(n)/n)$, converges to $(1-\gamma )^{-1}$ as $n\rightarrow
\infty$ and  $k(n)/n \rightarrow 0$.\\ 

 \noindent The Diop and Lo (2009) generalization of the Hill's estimator which corresponds to (\ref{hft})
was introduced in Diop and Lo (1994) and further studied in Diop and Lo (2009) where its
asymptotic normality was established for any $\gamma$, but only for $\tau >1/2$. It is not possible to
obtain the asymptotic distribution of the Diop and Lo estimator for $\tau \leq 1/2$
via Hungarian Gaussian approximation that was utilized then. Recently, the statistic obtained by dividing the expression in (\ref{hf1}) by $f(k)$, which generalizes (\ref{hft}) for $f(j)=j^{\tau}$, has been extensively studied for the Frechet and Gumbel distributions by Deme et al.(2012). When dealing with the family of functions $f(j)=j^{\tau}$, parameterized by $\tau>0$, we will use the notation
$f_{\tau}(j)=j^{\tau}$. In the aforementioned paper, Deme et al. (2012) proved that (\ref{hf1}) has a limiting
Gaussian process when $A(2,f)=: \sum_{j=1}^{+\infty }f(j)^{2}/j^{2}=\infty $ and
\begin{equation*}
B_{n}(f)=: \frac{\max \{f(j)/j,1\leq j\leq k\}}{\sqrt{\sum_{j=1}^{k}f(j)^{2}/j^2}}
\end{equation*}

\noindent converges to $0$ as $n\rightarrow \infty$. However, when $A(2,f)<+\infty$, (\ref{hf1}) has a non Gaussian
limiting process. Applying these results for the class of functions $f_{\tau }(\cdot)$, we get that for
$\gamma > 0$, the asymptotic law of (\ref{hft})  is Gaussian for $\tau \geq 1/2$ and non-Gaussian for $0<\tau < 1/2$. Coupling those observations with the results of Diop and Lo (2009), we can
easily conclude that the limiting distribution of (\ref{hft}) is known except for the Weibull domain and
for $0<\tau \leq 1/2$.\\

\noindent The problem of finding the asymptotic distribution of (\ref{hft}) in the Weibull domain when $0<\tau \leq 1/2$
has not been investigated. In this manuscript, we take a supermartingale approach to obtain the
limiting distribution of the generalized Hill statistics given in (\ref{hft}) in the said domain. The
supermartingale approach was motivated by the fact that under condition $A(2,f)< \infty$, (\ref{hf1}),
when properly centered and scaled can be written as a sum of dependent random variables.
With that approach and a complete characterization of the process
\begin{equation}
\sum_{j=1}^{k-1}(f(j)-f(j-1))\left[\exp (-\gamma \sum_{h=j}^{k-1}E_{h}/h)-
E\exp (-\gamma\sum_{h=j}^{k-1}E_{h}/h)\right],
\end{equation}

\noindent we were able  to find the asymptotic distribution of (\ref{hf1}) and (\ref{hft})
for $0 < \tau < 1/2$.\\

\noindent The manuscript will proceed as follow. In Section \ref{sec2}, we
study a special process based on a sequence of iid unit exponential
rvs whose limiting distribution will be found via martingale techniques.
Section \ref{sec3} is devoted to the application of the results of section \ref{sec2} to our
problem. The results of a simulation study are presented in section \ref{sec4} along with some results
pertaining to statistical tests in the extreme value domains. The manuscript will conclude with a discussion section and an appendix.

\section{A supermartingale tool}\label{sec2}
Let $E_{1},E_{2},...$ be iid unit exponentially distributed rvs. Let $k\geq 1$. Let also $\gamma>0$ be positive number that is constant throughout this paper. Define the increments of $f$ by $\Delta f(j)=f(j)-f(j-1)$ for $j=1,2,...$. Let also the sequence
$W_k(f)$ be given by the sequence
\begin{equation}
W_{k}(f)=\sum_{j=1}^{k}\Delta f(j)\left(\exp \left(-\gamma \sum_{h=j}^{k}E_{h}/h\right)-%
E\exp \left(-\gamma \sum_{h=j}^{k}E_{h}/h\right)\right),  \label{formdep}
\end{equation}

\noindent for $k\in\aleph$. Consider the filtration $\mathcal{F}_{k}=\sigma (E_{1},...,E_{k}),k\geq 1$
and observe that the sequence ($W_{k}$)$_{k\geq 1}$ is adapted to $(
\mathcal{F}_{k})_{k\geq 1}.$ We first introduce some intermediate
results that will prove to be crucial for the proof of our main results. The next two
ones pertain to the martingale property of $W_{k}(f)$ and its limiting distribution.
\begin{theorem} \label{theo1} The sequence $(W_{k}(f))_{k\geq 1}$ is a supermatingale with respect to $%
\mathcal{F}_{k}.$ Furthermore, it converges almost-surely (\textit{a.s}) to
a random variable $W_{\infty }(f)$ with finite expectation whenever
\begin{equation}
\limsup_{k\rightarrow +\infty }k^{-\gamma }\sum_{j=L}^{k-1}\Delta f%
(j)j^{\gamma -1/2}<+\infty  \tag{K1}
\end{equation}
holds.
\end{theorem}

\begin{corollary}
\label{cor1} For $f(j)=f_{\tau }(j)=j^{\tau }$, $0<\tau <1/2$, $W_{k}(f_{\tau})$
converges almost surely to a finite expectation random variable $W^{*}_{\infty }(\tau).$
\end{corollary}

\noindent \textbf{Remark} Actually, the random variable $W^{*}_{\infty}(\tau)$ given in Corollary \ref{cor1} is exactly $W^{}_{\infty }(f_{\tau})$  of Theorem \ref{theo1}, where the function $f_{\tau}$ is defined in Corollary \ref{cor1}. To make the notations simple, we will simply write 
$W^{*}_{\infty}(\tau)=W_{\infty}(\tau)$, when the argument is a positive number.\\
 
\noindent \bt{Proof of theorem \ref {theo1}:} Let
\begin{equation}
S_{j,k}=\exp \left(-\gamma \sum_{h=j}^{k}E_{h}/h\right)  \label{defs}
\end{equation}
\noindent and $V_{k}=\exp (-\gamma E_{k}/k)$ for $k\geq 1.$ We have for $k\geq 1$%
\begin{equation*}
W_{k+1}=\sum_{j=1}^{k}\Delta f(j)(S_{j,k}V_{k+1}-\mbox{E}\exp
(S_{j,k}V_{k+1}))+\Delta f(k+1)(V_{k+1}-\mbox{E}\exp (V_{k+1})).
\end{equation*}

\noindent Observe that : (1) $V_{k+1}$ and $S_{j,k}$ are
independent for $1\leq j\leq k$, (2) $V_{k+1}$ is independent of $%
\mathcal{F}_{k}$ and (3) an application of moment generating function
for unit exponential distribution yields $E(V_{k+1})=(1+\gamma
/(k+1))^{-1}$.\\

\noindent From now on, the notation $\ell(k,\gamma)=(1+\gamma/(k+1))^{-1}$ is used to make the formulas simple. The three previous observations, when combined together, lead to
$\mbox{E}((V_{k+1}-\mbox{E}\exp (V_{k+1}))|\mathcal{F}_{k})=0$ and
\begin{equation*}
\mbox{E}(W_{k+1}|\mathcal{F}_{k})=\mbox{E}(V_{k+1})\sum_{j=1}^{k+1}\Delta f%
(j)(S_{j,k}-\mbox{E}\exp (S_{j,k}))=\ell(k+1,\gamma)W_{k}.
\end{equation*}

\noindent Since the function $\ell(k,\gamma)$ is increasing in $k$, we obtain%
\begin{equation*}
\mbox{E}\left(\frac{W_{k+1}}{\ell(k+1,\gamma)}|\mathcal{F}_{k}\right)=\frac{W_{k}}{\ell(k,\gamma)}\times \frac{\ell(k,\gamma)}{\ell(k+1,\gamma)}\leq \frac{W_{k}}{\ell(k,\gamma)}.
\end{equation*}

\noindent Therefore, $W_{k}/\ell(k,\gamma)$ is a supermartingale. A
sufficient condition of a.s. convergence of $W_{k}/\ell(k,\gamma)$, as $k\rightarrow+\infty$ ($\gamma$ is fixed) to a 
finite expectation \textit{rv} is
\begin{equation*}
\limsup_{k\rightarrow +\infty }\mbox{E}(\left\vert W_{k}\right\vert
)< \infty,
\end{equation*}

\noindent since $\ell(k,\gamma)$ converges to unity when $k\rightarrow
\infty$, $\gamma$ being fixed.. Next, by denoting $s_{j,k}=E(S_{j,k})$, $1\leq j \leq k, k\geq 1$ and by using Cauchy-Schwarz and Minkowski inequalities, we
have
\begin{equation*}
E\left\vert W_{k}\right\vert \leq (\mbox{E}(W_{k}^2))^{1/2}=\left\Vert
W_{k}\right\Vert _{2}\leq \sum_{j=1}^{k}\left\Vert \Delta f(j)(S_{j,k}-s_{j,k})\right\Vert _{2}
\end{equation*}

\begin{equation*}
\leq \sum_{j=1}^{k-1}\Delta f(j)(\mbox{Var}(S_{j,k}))^{1/2}.
\end{equation*}

\noindent In the appendix, we provide moments computations of the
$S_{j,k}$'s, especially concerning their expectations, variances and covariances. These
computations are based on integral calculations given in
subsection \ref{subsec52}. By Formula (\ref{valueex}), $\mbox{Var}(S_{j,k})$ is
bounded by unity for any $1\leq j \leq k$. Combining this with (\ref{varsj}) and fixing $\varepsilon$ such that $0<\varepsilon \leq 1$, we get for $L$
large enough and $k-1>L$
\begin{eqnarray}
\mbox{E}\left\vert W_{k}\right\vert &\leq &\sum_{j=1}^{L}\Delta f%
(j)(\mbox{Var}(S_{j,k}))^{1/2}\nonumber\\&&+\frac{(2(1+\varepsilon )(\varepsilon +1/2))^{1/2}\gamma }{%
k^{\gamma }}\sum_{j=L+1}^{k-1}\Delta{f}(j)j^{\gamma -1/2}\nonumber\\
&\leq &\sum_{j=1}^{L}\Delta f(j)+\frac{4\gamma }{k^{\gamma }}%
\sum_{j=L+1}^{k-1}\Delta f(j)j^{\gamma -1/2}.\label{profth}
\end{eqnarray}%

\noindent Since the first term in the right hand side of (\ref{profth}) is bounded for a fixed $L$, we see that
the supremum limit of $\mbox{E}\left\vert W_{k}\right\vert $ is finite whenever (K1)
holds. This proves the theorem.\\

\bigskip \noindent \bt{Proof of Corollary:} Observe that for
large values of $j$, $\Delta f(j)\sim \tau j^{\tau -1}$ and Condition (K1)
is equivalent to the boundedness of
\begin{equation}
k^{-\gamma }\sum_{j=L}^{k-1}j^{\tau +\gamma -3/2}.\label{cd}
\end{equation}

\noindent Now,  let $0<\tau \leq 1/2$ and consider
the four possible cases: (i) $\tau +\gamma
-3/2=-1; (ii) $ $\tau +\gamma -3/2<-1$; (iii) $\tau
+\gamma -3/2=0$ and (iv ) $\tau +\gamma -3/2>0.$ Using (\ref{fgintegn}), we get
that (\ref{cd}) is less than $k^{-\gamma }(\log (k-1)-\log L)+(k-1)^{-1} $
for (i) and is less than $k^{-\gamma }((k-1)^{\gamma +(\tau -1/2)}-L^{\gamma
+(\tau -1/2)})/(\gamma +\tau -1/2)+(k-1)^{\gamma +(\tau -3/2)}$ for (ii). 
In the third case, (\ref{cd}) is exactly $k^{-\gamma
}(k-1-L)=(k^{(\tau -1/2)}-(1+L)k^{-(3/2-\tau )}.$ Finally, applying (\ref
{fgintegp}) to the last case, (\ref{cd}) is found to be less than
$$
k^{-\gamma }((k-1)^{\gamma +(\tau -1/2)}-L^{\gamma
+(\tau -1/2)})/(\gamma +\tau -1/2)+L^{\gamma +(\tau -3/2)}.
$$

\noindent In all these four cases, we get that $\lim \sup_{k\rightarrow
+\infty }\ k^{-\gamma }\sum_{j=L}^{k-1}j^{\tau +\gamma -3/2}<\infty$, and the statement
in the corollary is proven.\\

\section{Main Results}\label{sec3}

Let us give a brief statement about the problem of interest.\\

\noindent To make the notation compact, we set $k(n)\equiv k$. Recall that $G$
$\in D(G_{-\gamma})$ if and only if $F\in D(G_{-\gamma })$. As mentioned earlier,
this manuscript pertains to the properties of the leading part of
(\ref{hf1}) when $F\in D(G_{-\gamma})$ and  $0<\gamma<1/2$. We begin with the
special case of functions $F\in D(G_{-\gamma})$, that is
\begin{equation}
y_{0}-G^{-1}(1-u)=u^{\gamma}I\{u\in [0,1]\},  \label{fd}
\end{equation}

\noindent where $y_{0}$ is the upper endpoint of $G$. We use here the index $-\gamma <0$ instead of $\gamma <0$, hence condition (\ref{fd}) still holds. The next theorem characterizes the asymptotic distribution of $T_{n}(f)$ when Condition (K1) is satisfied.\\

\begin{theorem} \label{theo2} Let $X_{1},X_{2},...$ be a sequence of $iid$ rv's with common $\textit{df}$ $F$ such that (\ref{fd}) holds for $G(x)=F(e^x)$. Let $f(j)$ be an increasing function of $j \geq 0$, with $f(0)=0$ such that (K1) holds. For any $1\leq
k\leq n$, let

\begin{equation*}
A_{k,n}(f)=: f(k-1)-\sum_{j=1}^{k-1}(f(j)-f(j-1))\exp \left(
-\sum_{h=j}^{k-1}\log (1+\gamma /h)\right) .
\end{equation*}%

\noindent Then
\begin{equation*}
W_{k-1,n}^{\ast }(f):=A_{k,n}(f)-\frac{T_{n}(f)}{y_{0}-Y_{n-k+1,n}}
\end{equation*}

\noindent converges in distribution to the finite expectation random variable $W_{\infty }(f)$
defined in Theorem \ref{theo1}. Furthermore, if $f(j)=f_{\tau }(j)=j^{\tau }$ for $
0<\tau \leq 1/2,$ then $W_{k-1,n}^{\ast }(f_{\tau })$ converges in distribution
to $W_{\infty }(f_{\tau})=W^{*}_{\infty }(\tau)$ defined in Corollary \ref{cor1}.
\end{theorem}

\bigskip  \noindent \bt{Proof of Theorem \protect\ref{theo2}:}
The proof uses the classical representation of the $Y_{j}=\log X_{j} $
associated with the \textit{df} $G(x)=F(e^{x})$ through a sequence of
independent standard uniform rv's $U_{1},U_{2},...$ , that is%
\begin{equation*}
\{Y_{j},j\geq 1\}=_{d}\{G^{-1}(1-U_{j}),j\geq 1\}
\end{equation*}

\noindent and then
\begin{equation*}
\left\{ \{Y_{n-j+1,,n},1\leq j\leq n\},n\geq 1\right\} =_{d}\left\{
\{G^{-1}(1-U_{j,n}),1\leq j\leq n\},n\geq 1\right\} .
\end{equation*}

\noindent This gives
\begin{eqnarray*}
\frac{T_{n}(f)}{y_{0}-Y_{n-k+1,n}}&=&\sum_{j=1}^{k}f(j) \frac{ \log
X_{n-j+1,n}-\log X_{n-j,n}}{y_{0}-Y_{n-k+1,n}}\nonumber\\
&=&\sum_{j=1}^{k}f(j)\frac{(y_{0}-\log X_{n-j,n})-(y_{0}-\log
X_{n-j+1,n})}{y_{0}-Y_{n-k+1,n}}\nonumber\\
&=_d &\sum_{j=1}^{j=k}f(j)((U_{j+1,n}/U_{k,n})^{\gamma
}-(U_{j,n}/U_{k,n})^{\gamma }).
\end{eqnarray*}

\noindent We have for $1\leq j\leq k-1$
\begin{eqnarray*}
\left[ \frac{U_{j,n}}{U_{k,n}}\right]^\gamma
&=&\prod\limits_{h=j}^{k-1}\left[\frac{U_{h,n}}{U_{h+1,n}}\right]^\gamma\nonumber\\
&=&\exp \left(-\gamma \sum_{h=j}^{k-1}\frac{1}{h}\log \left( \frac{U_{h+1,n}}{U_{h,n}}\right)^h\right)\nonumber\\
&\equiv &\exp \left(-\gamma \sum_{h=j}^{k-1}E_{h}^{(n)}/h\right).
\end{eqnarray*}

\noindent By the Malmquist representation (see proposition 31 in \cite{wc-srv-ang}, p. 112 or Shorack and Wellner (1986), p. 336), the
rvs $\mbox{E}_{h}^{(n)}$ for $1\leq h\leq n$ are independent unit exponential. From that observation,
it follows that
\begin{equation*}
\begin{split}
\frac{T_{n}(f)}{y_{0}-Y_{n-k+1,n}}
&=\sum_{j=1}^{k}f(j)\left[ \exp \left(-\gamma
\sum_{h=j+1}^{k-1}\mbox{E}_{h}^{(n)}/h\right)-\exp \left(-\gamma
\sum_{h=j}^{k-1}\mbox{E}_{h}^{(n)}/h\right)\right].
\end{split}
\end{equation*}

\noindent Some algebraic manipulations yield
\begin{equation*}
\frac{T_{n}(f)}{y_{0}-Y_{n-k+1,n}}=f(k-1)-\biggl(\sum_{j=1}^{j=k-1}(f(j)-f(j-1))%
\exp (-\gamma \sum_{h=j}^{k-1}\mbox{E}_{h}^{(n)}/h)\biggr).
\end{equation*}

\noindent Set
\begin{equation*}
S_{j,k,n}=\exp \left(-\gamma \sum_{h=j}^{k-1}\mbox{E}_{h}^{(n)}/h\right)
\end{equation*}

\noindent and observe that for each $n\geq 1,S_{j,k,n}$ and $S_{j,k}$, which are defined in (\ref{defs}), have the same distribution. Let
\begin{equation*}
s_{j,k,n}=\mbox{E}(S_{j,k,n})=\exp \left(-\sum_{h=j}^{k-1}\log (1+\gamma /h)\right).
\end{equation*}

\noindent Then we have
\begin{equation}
A_{k,n}(f)=f(k-1)-\sum_{j=1}^{k-1}\Delta{f}(j)s_{j,k,n}.
\label{esperance}
\end{equation}

\noindent This yields
\begin{equation}
W_{k-1,n}^{\ast }(f)=\sum_{j=1}^{k-1}\Delta{f}(j)(S_{j,k,n}-s_{j,k,n}).
\label{formfin}
\end{equation}

\noindent Next, observe that for any $n\geq 1$, $W_{k(n)-1,n}^{\ast }(f)=_{d}W_{k(n)-1}$.
Therefore, $W_{k(n)-1,n}^{\ast }(f)$ converges in distribution to $W_{\infty }(f)$
whenever $W_{k(n)-1}$ converges almost surely to $W_{\infty }(f)$,  which
completes the proof of the theorem.

\section{Application to extreme value theory}\label{sec4}
\subsection{Asymptotic results in the Weibull case}
In this section, we are interested in the particular case of the Weibull distribution which obtained
for $0<\tau <1/2$. For the  general case, we have the following Karamata representation when $F$ is Weibull 
distributed with parameter $\gamma >0$, $x_{0}(F)<\infty $
\begin{equation}
\log x_{0}-F^{-1}(1-u)=c(1+p(u))u^{\gamma }\exp \left(\int_{u}^{1}b(t)t^{-1}dt\right),
\label{repweib}
\end{equation}

\noindent where $(p(u),b(u))\rightarrow (0,0)$ as $u\rightarrow 0$. Theorem 2 dealt with the special case of 
$p(u)=b(u)=0$.  In this section, we provide statistical tests for this special case
and further consider perturbation models that arise with special cases of $b(u)$.\\

\subsection{Critical points of the d.f. of $W_{\infty}(f)$}\label{subsec42}
\indent We use computer-based methods for approximating the law of $W_{\infty}(f)$.
Simulation studies show that the empirical d.f. based of $B_0=1000$
replications are very stable from $k=2000$.

\noindent We proceed as follows. Fix $\tau$ in $0<\tau <1/2,$ $\gamma >0$ and $%
k\geq 2000$. At each step, $B$ runs from 1 to $B_0=1000$. We generate standard
exponential samples $E_{1}(B),...,E_{k}(B)$ and compute $W_{k}^{\ast }$
denoted by $W_{k}^{\ast }(B)$. We finally consider the empirical d.f.
denoted by $G_{k}$, based on $W_{k}^{\ast }(1),\ldots,W_{k}^{\ast }(B_0)$. Since
$G_{k}$ is stable in the sense that it does not significantly change from $%
k=2000$, we approximate the d.f. $G_{\infty}$ of $W_{\infty}(f_{\tau })$
by $G_{k}$ for large $k$. As an example, we give an illustration in Figure \ref{figure1} the d.f.
$G_{k}$ for $k$ in $\{250, 500, 750, 1000, 2000, 5000\}$;  $\gamma =1$ and
$\tau=1/4$. Here for instance, we infer that the support of $G_{\infty }$ is
$[-0.5,0.5]$. Overall, the figures clearly establish stability and
support our proposal. Users who are interested in using the method provided in this paper may require executable files from the authors,
the computation of $P(W_{\infty}(f_{\tau })\leq x)=G_{\infty
}(x)$ and $P(\left\vert W_\infty(f_{\tau })\right\vert \leq \left\vert x\right\vert
)=G_{\infty}(\left\vert x\right\vert )-G_{\infty }(-\left\vert x\right\vert
) $ for $x\in \Re$.

\begin{figure}[htbp]
\centering
\caption{Illustration of the distribution functions of $W_{k,n}(1/4)$ for
different values of k}
\label{figure1}
\end{figure}

\begin{figure} 
	\centering
		\includegraphics{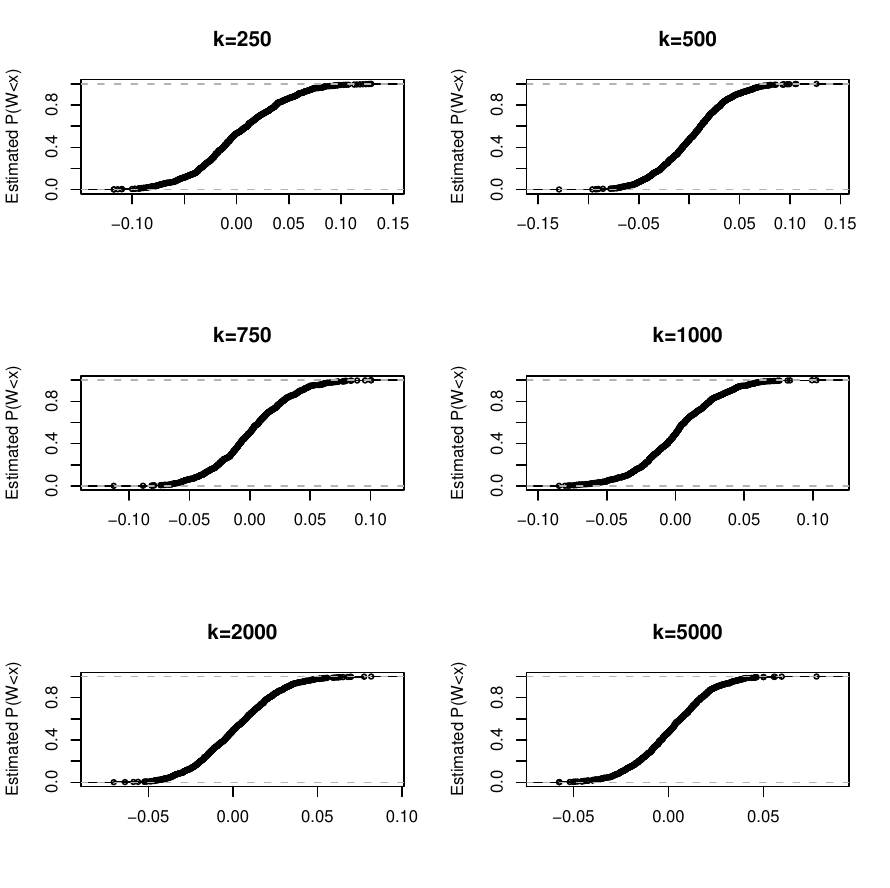}
	\label{fig:figure1}
\end{figure}

\begin{table}[tbp]
\centering
\begin{tabular}{llll}
\hline
Models & Quantile functions & $T_{n}^{\ast }(f_{\tau })$ & P-values \\ \hline
Weibull 1 & $F^{-1}(1-u)=\exp (1-u^{\gamma })$ & 3.16 & 67.4\% \\ \hline
Weibull 2 & $F^{-1}(1-u)=\exp (1-u^{\gamma }(1+u^{9}))$ & 0.0367 & 38.9\% \\
\hline
Weibull 3 & $F^{-1}(1-u)=\exp (1-u^{\gamma }(1+u^{8}))$ & 0.048 & 27.3\% \\ \hline
Weibull 4 & $F^{-1}(1-u)=\exp (1-u^{\gamma }(1+u^{7}))$ & 3.063 & 13.2\% \\ \hline
Weibull 5 & $F^{-1}(1-u)=\exp (1-u^{\gamma }(1+u^{6}))$ & 3.0725 & 10.4\% \\ \hline
Weibull 6 & $F^{-1}(1-u)=\exp (1-u^{\gamma }(1+u^{5}))$ & 3.097 & 2\% \\ \hline
Weibull 7 & $F^{-1}(1-u)=\exp (1-u^{\gamma }(1+u^{4}))$ & 3.17 & 0\% \\ \hline
Standard Exponential & $F^{-1}(1-u)=-\log u$ & 3.77 & 0\% \\ \hline
Pareto & $F^{-1}(1-u)=u^{-1}$ & 19.755 & 0\% \\ \hline
\end{tabular}%
\caption{Statistical tests for four models using the law of $W_{\infty}(1/4)$}
\label{table1}
\end{table}
\subsection{Statistical tests}\label{subsec43}
\indent In this subsection, we show how $G_{\infty }$\ may be used to test the null hypothesis
that $F\in D(G_{-\gamma }).$ The following approximation is used in the sequel.
\begin{equation*}
T_{n}^{\ast }(f)=T_{n}(f)/(y_{0}-\log X_{n-k+1,n})\approx T_{n}(f)/(\log
X_{n,n}-\log X_{n-k+1,n}).
\end{equation*}

\noindent For testing the null hypothesis that  $F\in D(G_{-\gamma})$, we compute the p-values for the seven models with $\gamma=1$, as described in Table \ref{table1}. The first corresponds to the \textit{pure model} for which $p(u)=b(u)=0, u\in (0,1)$. In the remaining six others, a shift of order $(1+u^q)$ is included in order to assess the impact of the perturbation, especially that of exponent $q$.
We use $n=300$ and $k=200$. The conclusions that transpired are as follow: (i) the pure model is accepted with a large p-value around $68\%$, (ii) the models Weibull 2 up to Weibull 6, which correspond to shift parameters $q \in\{6, 7, 8, 9\}$, are accepted with $p$-values at least greater than $10.4\%$; and (iii) the Weibull 6 and Weibull 7 models, corresponding to shift parameters $q=4$ and $q=5$, and the exponential and Pareto models are rejected. This is reasonable since, as we pointed out earlier, the convergence depends on the functions $b$ and $p$ in $(\ref{repweib})$ that are given by $p(u)=0$ and $b(u)=-qu^q(1+u^q)^{-1}$ and $c=1$.

\section{Remarks and Discussion} \label{sec6}\label{subsec41}
\indent We observe that for the Weibull simple case, the
law of the functional Hill process is found for $0<\tau <1/2.$ For the
general case, we have the following Karamata representation when $F$ is in
the Weibull case of parameter $\gamma >0$ : $x_{0}(F)<\infty $ and%
\begin{equation}
\log x_{0}-F^{-1}(1-u)=c(1+p(u))u^{\gamma }\exp \left(\int_{u}^{1}b(t)t^{-1}dt\right),
\label{repweib}
\end{equation}

\noindent where $(p(u),b(u))\rightarrow (0,0)$ as $u\rightarrow 0$. In an upcoming paper we will study the dependence of the
results on the auxiliary functions $p$ and $b$ and we also determine general conditions on $b$ and $p$ under which $T_{n}^{\ast
}(f)= T_{n}(f)/(y_{0}-\log X_{n-k+1,n})$ behaves as $W_{k,n}^{\ast }$ as in the present case.
Nevertheless, we will include in the statistical tests some models with
specific forms of $b(\cdot)$ as shown in Table 1 and used in Subsection \ref{subsec43}.

\section{Appendix}\label{sec5}
This section is devoted to the computations of the moments of
\begin{equation*}
S_{j,k}=\exp\left(-\gamma \sum_{h=j}^{k-1}E_{h}/h\right)
\end{equation*}

\noindent and to their approximations for large values of $j$, where the $E_{h}^{\prime }s$ are
independent unit exponential r.v. We begin by giving a
particular and useful tool for the expansion of the logarithm function. Let $\varepsilon >0$ be fixed.
There exists $u_{0}>0$ such that%
\begin{equation}
0<u<u_{0},\text{ }\log (1+u)=u+\theta (\varepsilon ,u)u^{2}, \label{expanInstrument}
\end{equation}

\noindent where $\theta (\varepsilon ,u)\in \lbrack -\varepsilon -1/2,\varepsilon
-1/2]\equiv A(\varepsilon )=[a_{1}(\varepsilon ),a_{2}(\varepsilon )].$ For
any integer $m\geq 1,$ let $J_{0}(m)$ be a positive integer such that $J_{0}(m)\geq \gamma
/(mu_{0})$. Then, we have
\begin{equation}
j\geq J_{0}(m)\Longrightarrow \log (1+\gamma /j)=\frac{j}{\gamma}+\theta _{j}\left(\frac{j}{\gamma}\right)^{2}\text{
with }\theta _{j}\in A(\varepsilon). \label{nounMoment01}
\end{equation}

\noindent This expansion (\ref{expanInstrument}) will be instrumental in the computations below. All coming numbers of the form $\theta_j$ and  $\theta_{h}(r)$ stand as notations of the number $\theta(u,\varepsilon)$ in (\ref{expanInstrument}) depending of the form of $u$. In all cases, these numbers are in $A(\varepsilon)$.\\

\subsection{Moment estimation}\label{subsec51}
\subsubsection{Exact values}\label{subsubsec511}

For any integer $m\geq 1$, we have 

\begin{equation*}
E\left( S_{j,k}^{m}\right) =E\exp (-m\gamma
\sum_{h=j}^{k-1}E_{h}/h)=\prod\limits_{h=j}^{k-1}E\exp (-m\gamma
E_{h}/h)=\prod\limits_{h=j}^{k-1}(1+m\gamma /h)^{-1}
\end{equation*}

\begin{equation}
=\exp \left( -\sum_{h=j}^{k-1}\log
(1+m\gamma /h)\right) .  \label{momentgen}
\end{equation}

\noindent Now for $j\geq J_{0}(m)$,
\begin{eqnarray}
E\left( S_{j,k}^{m}\right) &=&\exp \left(-m\gamma \sum_{h=j}^{k-1}(1/h)-m^{2}\gamma ^{2}\sum_{h=j}^{k-1}\theta_{h}/h^{2}\right) \label{expect01}
\end{eqnarray}

\noindent Then for any $j$ and $k$, $\mbox{Var}(S_{j,k})$ is
\begin{equation}
\exp \left( -2\sum_{h=j}^{k-1}\log (1+2\gamma /h)\right) -\exp \left(-2\sum_{h=j}^{k-1}\log (1+\gamma /h)\right) \leq 1,  \label{valueex}
\end{equation}

\noindent since this is difference of two points in $[0,1]$. Later, we will need this notation,
\begin{eqnarray}
E\left( S_{j,k}^{m}\right) &=&\exp \left(-m\gamma \sum_{h=j}^{k-1}(1/h)-m^{2}\gamma ^{2}\sum_{h=j}^{k-1}\theta_{h}/h^{2}\right) \label{expect03}\\
&=& \exp \left(-m\gamma \sum_{h=j}^{k-1}(1/h)\right)B(1,j,m) \notag
\end{eqnarray}

\noindent where

$$
B(1,j,m)=\exp \left(-m^{2}\gamma ^{2}\sum_{h=j}^{k-1}\theta_{h}/h^{2}\right)
$$

\subsubsection{Approximate values for moments}\label{subsubsec512}

Combining Formula (\ref{nounMoment01}) above and Formula (\ref{integ2}) in the last subsection \ref{subsec52} of this Appendix, leads to 

\begin{eqnarray}
\left\vert m^{2}\gamma ^{2}\sum_{h=j}^{k-1}\theta _{h}/h^{2}\right\vert &\leq& \left\vert a_{1}(\varepsilon )\right\vert m^{2}\gamma \left(\frac{1}{j}-\frac{1}{k-1}-\frac{1}{j^{2}}\right) \label{expect02} \\
&\leq& \frac{\left\vert a_{2}(\varepsilon)\right\vert m^{2}\gamma }{j}, \notag
\end{eqnarray}

\noindent for $j\geq J_{0}(m)$.\\

\noindent Let $J_{1}(\varepsilon,m)$ be a positive integer such that $\frac{\left\vert a_{1}(\varepsilon )\right\vert m^{2}\gamma }{J_{1}(\varepsilon,m)}\leq \varepsilon$. Then we have
\begin{equation*}
j\geq J_{1}(\varepsilon ,m)\vee J_{0}(m)\Longrightarrow \exp \left(
-m^{2}\gamma ^{2}\sum_{h=j}^{k-1}\theta _{h}/h^{2}\right) \leq
e^{\varepsilon }.
\end{equation*}

\noindent From the definition of $B(1,j,m)$ in (\ref{expect03}) above and from (\ref{expect02}), we have for $j \geq J_{0}(m)$

$$
B(1,m, j)=1+O(\frac{\left\vert a_{1}(\varepsilon )\right\vert m^{2}\gamma}{j})=1+O(j^{-1}),
$$

\noindent for $m$ fixed, since $\gamma$ is fixed in all the text. Next, by denoting
\begin{equation}
B(2,m,j)=\exp\left (-m\gamma \left\{ \sum_{h=j}^{k-1}\frac{1}{h}-\log((k-1)/j)\right\} \right), \label{formulrB2MJ}
\end{equation}

\noindent we get by (\ref{integ1}) that 

\begin{equation*}
\exp (-(m\gamma)/j)\leq B(1,m,j)\leq \exp (-(m\gamma)/(k-1))
\end{equation*}

\noindent so that, since $1\leq j\leq k$ and since $\gamma$ is fixed, for $m$ fixed,

$$
B(1,m,j)=1+O(j^{-1}), \ \ \text{ as j increases indefinitely.}
$$

\noindent Now, since 

\begin{equation}
\normalsize
\exp \left(-m\gamma \sum_{h=j}^{k-1}(1/h)\right)=\left( \frac{j}{k-1}\right) ^{m\gamma
}\exp \left(-m\gamma \left\{ \sum_{h=j}^{k-1}\frac{1}{h}-\log ((k-1)/j)\right\}\right), \label{expect04}
\end{equation}

\noindent we have, from Formula (\ref{expect01}) in the last subsection of this appendix and Formula (\ref{formulrB2MJ}) above, that 

\begin{equation}
E\left( S_{j,k}^{m}\right) =\left( \frac{j}{k-1}\right) ^{m\gamma
}B(1,m,j)B(2,m,j),  \label{momentsj}
\end{equation}

\noindent with,  for $m$ fixed and for $j\geq J_{1}(\varepsilon ,m)\vee J_{0}(m)$, 

\begin{equation*}
B(1,m,j)=1 + O(j^{-1}) \text{ and }B(2,m,j)=1+O(j^{-1}).
\end{equation*}

\subsubsection{Approximate values for variances}\label{subsubsec513}
We have for $j>J_{0}(2)$%
\begin{equation*}
E\left( S_{j,k}^2\right) =\exp \left( -2\gamma
\sum_{h=j}^{k-1}(1/h)-4^{2}\gamma ^{2}\sum_{h=j}^{k-1}\theta
_{h}(1)/h^{2}\right) .
\end{equation*}

\noindent and for $j>J_{0}(1)$
\begin{equation*}
E\left( S_{j,k}\right)^2 = \exp  \left( \biggr( -\gamma\sum_{h=j}^{k-1}(1/h)-\gamma ^{2}\sum_{h=j}^{k-1}\theta _{h}(2)/h^{2} \biggr) \right)
^{2}
\end{equation*}%
\begin{equation*}
=\exp \left( -2\gamma \sum_{h=j}^{k-1}(1/h)-2\gamma
^{2}\sum_{h=j}^{k-1}\theta _{h}(2)/h^{2}\right) .
\end{equation*}

\noindent Thus
\begin{equation}
Var(S_{j}^{\ast })=\exp \left(2\gamma \sum_{h=j}^{k-1}1/h\right) \times V(2,2,j)\label{varNous01}
\end{equation}

\noindent where  

\begin{equation*} \large
V(2,2,j)=\left\{ \exp \left(-4^{2}\gamma ^{2}\sum_{h=j}^{k-1}\theta_{h}(1)/h^{2}\right)-\exp \left(-2\gamma ^{2}\sum_{h=j}^{k-1}(2\theta _{h}(1)-\theta_{h}(2))/h^{2}\right)\right\} . 
\end{equation*}

\noindent Now, the same technique used in Formula (\ref{expect04}) above proves that for $m=2$,

$$
\exp \left(2\gamma \sum_{h=j}^{k-1}1/h\right)=\left(\frac{j}{k-1}\right)^{2\gamma} V(1,2,j),
$$

\noindent with $V(1,2,j)=1+O(j^{-1})$ as $j \rightarrow +\infty$.\\

\noindent Let us now handle $V(2,2,j)$. \noindent Since $x=4^{2}\gamma ^{2}\sum_{h=j}^{k-1}\theta _{h}(1)/h^{2}$ and $%
y=2\gamma ^{2}\sum_{h=j}^{k-1}\theta _{h}(2)/h^{2}$ are both nonnegative, we
have $\left\vert e^{-x}-e^{-y}\right\vert $ $\leq \left\vert x-y\right\vert$.
Thus
\begin{eqnarray}
0 &\leq &\exp \left(-4^{2}\gamma ^{2}\sum_{h=j}^{k-1}\theta _{h}(1)/h^{2}\right)-\exp
\left(-2\gamma ^{2}\sum_{h=j}^{k-1}\theta _{h}(2))/h^{2}\right)\nonumber\\
&\leq & 2\gamma ^{2}\sum_{h=j}^{k-1}\left\vert 2\theta _{h}(1)-\theta
_{h}(2)\right\vert /h^{2}\leq \frac{2\gamma ^{2}\left\vert a_{2}(\varepsilon
)\right\vert }{j},
\end{eqnarray}

\noindent by (\ref{integ2}).  Therefore
\begin{equation}
Var(S_{j,k})=\left( \frac{j}{k-1}\right) ^{2\gamma }V(1,2,j)V(2,2,j) \label{varsj}
\end{equation}

\noindent with

\begin{equation*}
\left\vert V(1,2,j)\right\vert =1+O(j^{-1})\text{ and } V(2,2,j)=1+O(j^{-1}).
\end{equation*}

\subsubsection{Covariance approximate values}\label{subsubsec514}
Let $\ell >1$ and consider $\sigma _{j,j+\ell }=\mbox{cov}(S_{j+\ell ,k},S_{j,k}).$
We have
\begin{eqnarray}
E\left( S_{j,k}\right)& =&\exp\left(\sum_{h=j}^{k-1}-\log (1+\gamma /h)\right)\nonumber\\
&=&\exp \left(\sum_{h=j}^{j+\ell -1}-\log (1+\gamma /h)\right)\exp\left(\sum_{h=j+l
}^{k-1}-\log (1+\gamma /h)\right)\nonumber\\
&=&E\left( S_{j+\ell ,k}{}\right) \exp \left(\sum_{h=j}^{j+\ell -1}-\log (1+\gamma
/h)\right).
\end{eqnarray}

\noindent Further,
\begin{eqnarray}
S_{j,k}S_{j+\ell ,k}{}&=&\exp \left(-\gamma \sum_{h=j}^{k-1}E_{h}/h\right)\exp\left(-\gamma
\sum_{h=j+\ell }^{k-1}E_{h}/h\right)\nonumber\\
&=&\exp \left(-\gamma \sum_{h=j}^{j+\ell -1}E_{h}/h-\gamma \sum_{h=j+\ell
}^{k-1}E_{h}/h\right)\exp \left(-\gamma \sum_{h=j+\ell }^{k-1}E_{h}/h\right)\nonumber\\
&=&\exp \left(-2\gamma \sum_{h=j+\ell }^{k-1}E_{h}/h\right)\exp \left(-\gamma
\sum_{h=j}^{j+\ell -1}E_{h}/h\right)\nonumber\\
&=& S_{j+\ell ,k}^{2}{}\exp \left(-\gamma
\sum_{h=j}^{j+\ell -1}E_{h}/h\right).
\end{eqnarray}

\noindent Hence

\begin{equation*}
E(S_{j,k}S_{j+\ell ,k}{}{})=E\left( S_{j+\ell ,k}\right) ^{2}\exp
\left(\sum_{h=j}^{j+\ell -1}-\log (1+\gamma /h)\right).
\end{equation*}%

\noindent For $j\geq J_{0}(1)\vee J_{0}(2),$%
\begin{eqnarray}
\mbox{cov}(S_{j,k}S_{j+\ell ,k}{}{})&=&\mbox{Var}(S_{j+\ell ,k})\exp \left(\sum_{h=j}^{j+\ell
-1}-\log (1+\gamma /h)\right)\nonumber\\
&=&\mbox{Var}(S_{j+\ell ,k}{})\exp \left(-\gamma
\sum_{h=j}^{j+\ell -1}1/h-\gamma ^{2}\sum_{h=j}^{j+\ell -1}\theta _{h}/h^{2}\right)\nonumber\\
&=& \mbox{Var}(S_{j+\ell ,k}{})\left( \frac{j}{j+\ell -1}\right) ^{\gamma
}(1+O(j^{-1}))  \label{covarsj}
\end{eqnarray}

\subsection{Integrals computations}\label{subsec52}

Let $b\geq 1$. By comparing the area under the curve of $f(x)=x^{-b}$ going from $j$ to $k-1$
and that of the rectangles based on the intervals
$[h,h+1],$ $h=1,..,k-2,$ we obtain%
\begin{equation*}
\sum_{h=j+1}^{k-1}h^{-b}\leq \int_{j}^{k-1}x^{-b}dx\leq
\sum_{h=j}^{k-2}h^{-b},
\end{equation*}%
that is%
\begin{equation}
\int_{j}^{k-1}x^{-b}dx+(k-1)^{-b}\leq \sum_{h=j}^{k-1}h^{-b}\leq
\int_{j}^{k-1}x^{-b}dx+j^{-b}.  \label{fgintegn}
\end{equation}

\noindent Likely, comparing the area under the curve of $f(x)= x^{b}$
going from $j$ to $k-1$ and those of the rectangles based on the intervals $%
[h,h+1],$ $j=1,..,k-2,$ we also get
\begin{equation}
\int_{j}^{k-1}x^{b}dx+j^{b}\leq\sum_{h=j}^{k-1}h^{b}\leq
\int_{j}^{k-1}x^{b}dx+(k-1)^{b}.  \label{fgintegp}
\end{equation}
\noindent Next, for $b=1$ and $b=2$, (\ref{fgintegn}) yields%
\begin{equation}
\frac{1}{j}\leq\log ((k-1)/j)-\sum_{h=j}^{k-1}\frac{1}{h}\leq \frac{1}{k-1%
}.  \label{integ1}
\end{equation}

\noindent and
\begin{equation}
\frac{1}{j}-\frac{1}{k-1}-\frac{1}{j^{2}}\geq\sum_{h=j}^{k-1}h^{-2}\geq
\frac{1}{j}-\frac{1}{k-1}-\frac{1}{(k-1)^{2}},  \label{integ2}
\end{equation}

\noindent respectively. Combining both implications, we further get%
\begin{equation*}
\frac{1}{j^{2}}\leq\frac{1}{j}\left(1-\frac{j}{k-1}\right)-\sum_{h=j}^{k-1}h^{-2}%
\leq \frac{1}{(k-1)^{2}}.
\end{equation*}

\begin{acknowledgement}
The authors thank the unknown referee whose detailed and careful reading, and comments and suggestions significantly helped in making the paper better in all its aspects.
\end{acknowledgement}

\begin{acknowledgement}
The two first authors acknowledge support from the World Bank Excellence Center (CEA-MITIC) of Saint-Louis, Senegal, that is continuously funding their research activities starting 2014.
\end{acknowledgement}


\begin{thebibliography}{99}
\bibitem{bgt} Beirlant, J. Goegebeur, Y. and Teugels, J.(2004). \textit{%
Statistics of Extremes Theory and Applications}. Wiley. (MR2108013)

\bibitem{cdm} Cs\"{o}rg\H{o}, S., Deheuvels, P. and Mason, D. M.(1985).
Kernel estimates of the tail index of a distribution. Ann. Statist., 13,
1050-1077. (MR0803758)

\bibitem{daley} Daley, D.J.(1968). Stochastically monotone Markov chains, Z.
\textit{Wahrsch. theor. verw Gebiete}, 10,305-317.

\bibitem{demedioplo} D\`eme E., Lo G.S. and Diop, A.(2012). On the generalized Hill process for small parameters and applications. \textit{Journal of Statistical Theory and Application}, 11(2), pp. 397-418.  Doi : 10.2991/jsta.2013.12.1.3 

\bibitem{demelo} Lo G.S., D\`eme E.(2012). A Functional Generalized Hill Process and Its Uniform Theory.  \textit{International Journal of Statistics and Probability}, 1(2), pp. 250-268. Doi:10.5539/ijsp.v1n2p250

\bibitem{dioplo1} Diop, A. and Lo, G.S.(1994). Sur une caract\'erisation
statistique simple des extr\^emes. \textit{Afrika Mat.}, S\'erie 3, Vol.
(3), 81-95. (Zbl 0819.60052) (MR1430638) \newline

\bibitem{dioplo} Diop, A. and Lo G. S.(2009). Ratio of generalized Hill's
estimator and its asymptotic normality theory. \textit{Math. Method. Statist.%
}, 18(2), pp. 117-133. (MR2537361)

\bibitem{embrechts} Embrechts, P., K\H{u}ppelberg C. and Mikosh T. (1997).
Modelling extremal events for insurance and Finance. Springer Verlag.

\bibitem{galambos} Galambos, J.(1985). \textit{The Asymptotic Theory of
Extreme Order Statistics}. Wiley, New-York. (MR0489334)

\bibitem{gbw} Goegebeur, Y., Beirlant, J. and de Wet, T.(2010). Kernel estimators for the
second order parameter in extreme value statistics. \textit{J. Statist. Plann. Inference},
140 (9), 2632-2652. (MR2644084)

\bibitem{glw} Groeneboom, Lopuha\"{a}, H.P and Wolf, P.P.(2003). Kernel-type
estimator for the extreme values index. \textit{Ann. Statist.}, 31, (6), pp.
1956-1995.

\bibitem{dehaan} de Haan, L. and Ferreira A. (2006). \textit{Extreme Value
Theory: An Introduction}. Springer. (MR2234156)

\bibitem{karlin} Karlin, S. and Howard M.,Taylor (1975). \textit{A First
Course in Stochastic Processes}. Academic Press.

\bibitem{wc-srv-ang} Lo, G.S.(2016). Weak Convergence (IA). Sequences of random vectors. SPAS Books Series.(2016).
Doi : 10.16929/sbs/2016.0001. Arxiv : 1610.05415

\bibitem{loeve} Lo\`eve, M.(1977). \textit{Probability Theory I}.
Springer-Verlag. New-York.

\bibitem{resnick} Resnick, S.I.(1987). \textit{Extreme Values, Regular
Variation and Point Processes}. Springer-Verbag, New-York.

\bibitem{shwell} Shorack G.R. and Wellner J. A.(1986). \textit{Empirical
Processes with Applications to Statistics}. Wiley-Interscience, New-York.
(MR0838963)
\end{thebibliography}
\end{document}